\documentclass[twocolumn,amsmath,superscriptaddress]{revtex4}
\usepackage{amsmath}
\usepackage{mathrsfs}
\usepackage{graphicx}
\usepackage{multirow}

\newcommand{\vect}[1]{{\boldsymbol{\mathbf{#1}}}}
\newcommand{\jr}[1]{{\text{#1}}}

\begin{document}

\title{The Global Network of Optical Magnetometers for Exotic Physics (GNOME):\\ A novel scheme to search for physics beyond the Standard Model}

\author{S. Pustelny}
\email{pustelny@uj.edu.pl}
\affiliation{Institute of Physics, Jagiellonian University, Reymonta 4, 30-059 Krak\'ow, Poland}
\affiliation{Department of Physics, University of California at Berkeley, Berkeley CA, 94720-7300, USA}

\author{D. F. Jackson Kimball}
\affiliation{Department of Physics, California State University–East Bay, Hayward CA, 94542-3084, USA}

\author{C. Pankow}
\affiliation{Center for Gravitation, Cosmology, and Astrophysics, Department of Physics, University of Wisconsin-Milwaukee, 1900 E. Kenwood Blvd, Milwaukee WI, 53211, USA}

\author{M. P. Ledbetter\footnote{Current address: IOSence, Inc., 767 N Mary Ave, Sunnyvale CA, 94085-2909, USA}}
\affiliation{Department of Physics, University of California at Berkeley, Berkeley CA, 94720-7300, USA}

\author{P. Wlodarczyk}
\affiliation{Department of Electronics, AGH University of Science and Technology, Mickiewicza 30, 30-059 Krak\'ow, Poland}

\author{P. Wcislo}
\affiliation{Institute of Physics, Jagiellonian University, Reymonta 4, 30-059 Krak\'ow, Poland}
\affiliation{Institute of Physics, Faculty of Physics, Astronomy and Informatics, Nicolaus Copernicus University, Grudziadzka 5, 87-100 Toru\'n, Poland}

\author{M. Pospelov}
\affiliation{Department of Physics and Astronomy, University of Victoria, Victoria BC, V8P 1A1, Canada}
\affiliation{Perimeter Institute for Theoretical Physics, Waterloo ON, N2J 2W9, Canada}

\author{J. R. Smith}
\affiliation{Gravitational-Wave Physics and Astronomy Center, Department of Physics, California State University Fullerton, 800 N State College Blvd., Fullerton CA, 92831, USA}

\author{J. Read}
\affiliation{Gravitational-Wave Physics and Astronomy Center, Department of Physics, California State University Fullerton, 800 N State College Blvd., Fullerton CA, 92831, USA}

\author{W. Gawlik}
\affiliation{Institute of Physics, Jagiellonian University, Reymonta 4, 30-059 Krak\'ow, Poland}

\author{D. Budker}
\affiliation{Department of Physics, University of California at Berkeley, Berkeley CA, 94720-7300, USA}
\affiliation{Nuclear Science Division, Lawrence Berkeley National Laboratory, Berkeley CA, 94720, USA}

\date{\today}

\begin{abstract}
A novel experimental scheme enabling the investigation of transient exotic spin couplings is discussed. The scheme is based on synchronous measurements of optical-magnetometer signals from several devices operating in magnetically shielded environments in distant locations ($\gtrsim100$~km). Although signatures of such exotic couplings may be present in the signal from a single magnetometer, it would be challenging to distinguish them from noise. By analyzing the correlation between signals from multiple, geographically separated magnetometers, it is not only possible to identify the exotic transient but also to investigate its nature. The ability of the network to probe presently unconstrained physics beyond the Standard Model is examined by considering the spin coupling to stable topological defects (e.g., domain walls) of axion-like fields. In the spirit of this research, a brief ($\sim 2$~hours) demonstration experiment involving two magnetometers located in Krak\'ow and Berkeley ($\sim 9000$~km separation) is presented and discussion of the data-analysis approaches that may allow identification of transient signals is provided. The prospects of the network are outlined in the last part of the paper.
\end{abstract}
\maketitle

\section{Introduction}

Among all magnetometric techniques, optical magnetometry \cite{Budker2007Optical,Budker2013Optical} presently offers the possibility of the most sensitive magnetic-field measurements \cite{Dang2010Ultrasensitive}. Intrinsic sensitivity of optical magnetometers (OMAGs) to spin dynamics also enables investigation of other spin interactions, including non-magnetic ones (see Ref.~\cite{Budker2013Optical} and references therein). In particular, OMAGs can be applied to probe couplings between spins and hypothetical fields not predicted by the Standard Model. Such exotic fields are postulated by a variety of theories \cite{Venema1992Search,Wineland1991Search,Glenday2008Limits,Vesilakis2009Limits,Berglund1995New,Brown2010New,Ni2010Searches,
Youdin1996Limits,Pospelov2013Detecting,Ledbetter2013Constrains}. One manner in which they could manifest themselves on Earth is as transient events. A particular example would be transient coupling of spins to certain constituents of dark matter (DM) and dark energy (DE) \cite{Pospelov2013Detecting}.

Most experimental DM searches aim at direct detection of some variety of particles that feebly interact with ordinary baryonic matter, e.g., Weakly Interacting Massive Particles (WIMPs) or axions \cite{Bertone2005Particle}. Until now, however, all the searches have produced only upper limits on the interaction strength between DM and ordinary matter. Over the years alternative candidates for DM have been proposed. For example, if DM consists of light axions or axion-like particles, it behaves more like a coherent field than a collection of uncorrelated particles \cite{Preskill1983Cosmology,Dine1983}. In some theoretical scenarios, because the vacuum energy of the axion field is non-zero, the field oscillates at a specific frequency and hence it would not produce static effects on matter. Such scenarios might also generate stable topological defects \cite{Pospelov2013Detecting,Flambaum2009,Pospelov2008,Abbott1983,Kaloper2006,Carroll1998,Lue1999,Pospelov2009}, e.g., a domain structure \cite{Pospelov2013Detecting}. When the Earth crosses one of the domain walls (DWs) separating regions with different vacuum expectation values of the axion-like field, a torque can be exerted on leptonic or baryonic spins.  Such a DW-crossing event could lead to a transient signal detectable with modern state-of-the-art OMAGs \cite{Pospelov2013Detecting}. Based on astronomical constraints, however, one can show that wall-crossing events are rare and brief \cite{Pospelov2013Detecting}, so the major issue becomes separation of the  transient signals induced by the DW crossing from transient signals generated by environmental and technical noise. Reliable rejection of OMAG's transient signals due to other effects requires development of a new approach.

In this paper, the principles of a new technique for detecting transient signals of exotic origin using a global network of synchronized OMAGs (the Global Network of Optical Magnetometers for Exotic physics, GNOME) is demonstrated. Although the network may be used for detecting of a variety of transient interactions heralding physics beyond the Standard Model, here, for concreteness, the considerations are focused on the transient effects induced by crossing through the DWs of an axion-like field. It is demonstrated that the GNOME enables probing presently unconstrained parameters of the field.

The article is organized as follows. First, a general discussion of OMAGs is provided. The characteristics of OMAGs that are most relevant for detection of exotic transient events are emphasized. Next, a review of the theory of DWs of axion-like fields relevant to their detection by the GNOME is provided (Sec.~\ref{sec:Theory}). Section \ref{sec:Setup} discusses a demonstration experiment using synchronously detected signals of two OMAGs separated by $\sim9000$~km. These two magnetometers form the first sensors of the envisioned GNOME. The principles that form the basis for GNOME data analysis are outlined in Sec.~\ref{sec:Results}. Finally, prospects of the GNOME are discussed in Sec.~\ref{sec:Prospects} and conclusions are drawn in Sec.~\ref{sec:Conclusions}.

\section{Characteristics of OMAGs relevant for detecting transient effects\label{sec:Optical}}

The detection of transient events that weakly perturb atomic spins requires OMAGs with specific characteristics. In particular, a suitable device needs to have high enough sensitivity to detect the small changes in spin dynamics associated with heretofore undiscovered exotic physics. Moreover, its response to abrupt changes of spin behavior needs to be suitably fast not to distort or average out the signals, so that their time-domain signature can be reliably understood and compared between different GNOME sites. High sensitivity and high bandwidth, however, may not always be compatible with each other. Certain high-sensitivity OMAGs have characteristically slow spin relaxations, which lead to narrower bandwidths, whereas high bandwidth devices may have fast spin relaxation times that degrade sensitivity. Below, fundamental and technical limitations of OMAG sensitivity as well as factors determining the bandwidth of the devices are discussed. Discussion of other characteristics of the magnetometers, relevant for detection of transient signals, is also provided.

In OMAGs, the detection of magnetic fields occurs in a three-stage process. First, atoms are optically pumped; next, they evolve under the influence of external fields; finally, their quantum state is detected with light \footnote{The stages may either be separated in time or occur simultaneously.}. While this scheme allows for the most sensitive measurements of magnetic fields, it also sets a fundamental limit on the sensitivity of OMAGs. The limit results from the quantum nature of photons and atoms and the coupling between them. In general the fundamental limit of the sensitivity is $\delta B_f=\sqrt{\delta B_{at}^2 + \delta B_{ph}^2 + \delta B_{ba}^2}$, where $\delta B_{at}$ is the limit due to spin-projection noise (SPN), $\delta B_{ph}$ is the limit related to photon shot noise, and $\delta B_{ba}$ is the limit associated with backaction of the probe light on the atoms. The projection noise originates from the Heisenberg uncertainty principle $\delta F_i^2\delta F_j^2\geq|\langle[F_i,F_j]\rangle|^2/4=\hbar^2\langle F_k\rangle^2/4$, where $F_{i,j,k}$ are three components of the spin $\vect{F}$ and [,] denotes the commutator. When this relation becomes an equality, the SPN-limited magnetic-field sensitivity $\delta B_{at}$ may be written as \cite{Budker2013Optical}
\begin{equation}
    \delta B_{at}=\frac{\hbar}{g\mu_B}\sqrt{\frac{1}{N_{at}T_2\tau}},
    \label{eq:SensitivityPN}
\end{equation}
where $N_{at}$ is the total number of atoms involved in the light-atom interaction, $T_2$ is the transverse spin-relaxation time, $\tau$ is the duration of the measurement, $g$ is the Land\'e factor, $\mu_B$ is the Bohr magneton, $\hbar$ is the Planck constant, and $c$ is the speed of light \footnote{Although for a minimum uncertainty state, the sensitivity $\delta B_{at}$ scales as $1/\sqrt{N_{at}}$, for a quantum system with entanglement the scaling could be stronger ($\delta B_{at}\propto 1/N_{at}$) \cite{Geremia2003Quantum}. In principle, it should allow for a large improvement in magnetic-field sensitivity of OMAGs. Unfortunately, the entangled states are fragile and rarely improvement of the sensitivity $\delta B_{at}$ below Eq.~(\ref{eq:SensitivityPN}) is observed \cite{Auzinsh2004Can,Wasilewski2009Quantum,Sewell2012Magnetic}.}. Equation~(\ref{eq:SensitivityPN}) reveals two strategies to improve the sensitivity of OMAGs. The first consists in prolonging the transverse spin-relaxation time $T_2$, for example, by containing the atoms in a glass cell with antirelaxation coating that preserves spin polarization upon atomic collisions with cell walls or introducing a buffer gas with a low spin-depolarization cross-section into the cell to limit diffusion to the walls. The second approach relies on increasing the number of atoms $N_{at}$. Both approaches are used in OMAGs and in fact, herein experimental results obtained with OMAGs exploiting both methods are discussed (Sec.~\ref{sec:Results}).

The photon shot-noise-limited sensitivity $\delta B_{ph}$ is associated with the fluctuation in the number of photons in the light beam used for probing the spins. Due to Poissonian statistics of photons, the intensity and polarization-state of light can only be determined with a finite precision $\propto (\dot{N}_{ph}\tau)^{-1/2}$, where $\dot{N}_{ph}$ is a number of photons of the probe beam hitting detector per unit time. This sets a limit on the precision with which the spin state can be determined and hence the limit on the magnetometric sensitivity. It is important to note, however, that this contribution can be reduced by detuning the probe-light frequency away from the resonant optical transition and simultaneously increasing the light intensity. The photon shot-noise limited sensitivity $\delta B_{ph}$ improves due to the increase in $\dot{N}_{ph}$, and the probe light only weakly affects the medium while the state can still be efficiently determined (absorption on an isolated transition scales as $1/\Delta^2$ and dispersion as $1/\Delta$, where $\Delta$ is detuning). Hence, the contribution of the photon shot noise to the total magnetometric sensitivity $\delta B_f$ may be reduced so that $\delta B_{ph}\ll\delta B_{at}$.

The last source of fundamental noise comes from the Stark shift of energy levels induced by quantum fluctuations of light intensity and polarization (back action) \cite{Happer1967Effective}; fluctuations of energies of magnetic sublevels introduce uncertainty in spin precession and hence limit magnetometric sensitivity. Yet, there are means of reducing or eliminating backaction \cite{Novikova2001Compensation,Wasilewski2009Quantum,Vasilakis2011Stroboscopic}. For instance, for large detuning, when an atomic system can be treated as a spin-1/2 particle (the hyperfine structure is unresolved), the Stark shift scales inversely proportional to the square of the light detuning \cite{Jensen2009Cancelalation}, so that the backaction may be significantly reduced. In turn, under the optimized conditions, the fundamental sensitivity limit is determined by the atomic shot noise ($\delta B_f\approx\delta B_{at}$).

Typically, on top of fundamental noise, there is technical noise $\delta B_t$. For example, the noise may be induced by mechanical vibration of optical elements or air turbulence in the probe-beam optical path. Electronics used in light detection can also contribute to technical noise. With appropriate experimental measures, the influence of the noise on overall magnetometric sensitivity may be reduced but it cannot be completely eliminated, hence, in many cases, it is a significant (sometimes dominant) contribution to OMAG sensitivity.

A different source of noise originates from uncontrollable magnetic fields. Such fields generate a random response of OMAGs, which reduces their sensitivity to non-magnetic interactions affecting the atomic spins. In the case of OMAGs enclosed inside a magnetic shield, a common configuration for precision measurements, uncertainty in spin-dynamics measurements may be introduced by external magnetic fields penetrating into the shield \footnote{Note that the sensitivity of the magnetometer is at a level of 10$^{-15}$~T/$\sqrt{\text{Hz}}$ or better, while the Earth magnetic field is $\approx 4\times10^{-5}$~T. Hence even with a shielding factor of $>10^6$ for DC fields, the magnetometer is still strongly sensitive to the fluctuation of the external fields.}, thermal currents induced in the shield material, and instability of the current sources used for generating magnetic fields.  The noise may be reduced by application of active cancelation of the field outside the shield and/or by the use of low electric conductivity high magnetic susceptibility shielding materials \footnote{In order to limit magnetic-field noise due to Johnson currents in the shield, modern magnetic shields have the innermost layer made of ferrite. Although such material has significantly lower magnetic susceptibility than permaloys, in particular, $\mu$-metal, they have orders of magnitude larger resistivity, which suppresses thermal currents in the magnetic-shield layer placed closest to a vapor cell (see Ref.~\cite{Budker2013Optical} and references therein).}. Another approach is to employ comagnetometry techniques, where the magnetic field is measured by multiple species expected to have different couplings to the exotic fields, allowing subtraction or cancelation of magnetic field noise.

Although OMAGs do not have intrinsic $1/f$ noise, the presence of technical noise suggests an advantage of detection of optical signals at frequencies higher than the $1/f$-``knee''. This may be achieved either by modulation of the probe light, i.e., by application of intensity, frequency, or polarization modulation, and phase-sensitive detection of the signal, or by operation of the devices in non-zero magnetic fields $B\gg(T_2g\mu_B)^{-1}$. In the latter case, the output signal of the magnetometer is modulated at the Larmor frequency $\omega_L=g\mu_B B$  or a multiple thereof, which enables filtering of the low-frequency noise. To detect such higher-frequency signals, however, OMAGs with broad dynamic ranges are required.

Operation at non-zero magnetic fields raises another important issue in magnetic-field detection. Optical magnetometers enable either scalar measurements, where the device response depends on the magnitude of a magnetic field, or vector measurements, where it is determined by specific vector components of the field. However, even scalar magnetometers operating at non-zero magnetic fields become primarily sensitive to the field changes along the dominant component of the field; transverse components of the field add as second-order corrections to the total-field magnitude $B$. Moreover, modulation of the magnetic field in three spatial directions enables a scalar magnetometer to detect the three vector components of the field \cite{Alexandrov2004Three,Seltzer2004Unshielded}. There also exist techniques enabling conversion of a scalar magnetometer into a vector magnetometer without the necessity of applying a modulated magnetic field \cite{Pustelny2006Nonlinear}. The ability to determine not only a magnitude but also the direction of the spin-coupling field may have implications for the envisioned detection of transient effects due to exotic interactions.

Another characteristic of OMAGs, particularly important for the detection of transient signals, is bandwidth. For typical OMAGs (see Ref.~\cite{Budker2013Optical} and references therein), the response of the magnetometer to small field changes is equivalent to a response of a first-order low-pass filter with the time constant $T_2$ \cite{Wlodarczyk2012Modeling}. Hence the natural bandwidth of such OMAGs is given by $(2\pi T_2)^{-1}$, which for shorter measurement times, i.e., $\tau<T_2$, takes the form $(2\pi\tau)^{-1}$. OMAG bandwidth can be broadened by shortening $T_2$, which can be, for example, accomplished by increasing intensity of the probe light (power broadening). That increase of the magnetometer bandwidth often occurs at the cost of its sensitivity [Eq.~(\ref{eq:SensitivityPN})]. Therefore, optimized operation of OMAGs requires a compromise between the two quantities. It should be noted, however, that application of quantum nondemolition measurements enables sensitive magnetic-field measurements at high bandwidth \cite{Shah2010High,Koschorreck2010Sub}.

\begin{table*}[t]
    \caption{Various OMAG characteristics important for detecting transient signals due to exotic spin couplings. $\delta B_f$ and $\delta B_d$ are fundamentally limited and experimentally demonstrated OMAG sensitivities of magnetic-field measurements, respectively, $\delta E_f$ and $\delta E_d$ are corresponding sensitivities in energy units obtained by multiplication of the magnetometric sensitivity by appropriate atomic/molecular magnetic moments $\gamma$ ($\delta E_{\!f,d}=\gamma\ \delta B_{\!f,d}/\hbar$). The names of the magnetometers indicate the type of the device and Hg EDM stands for the magnetometer used in the experiment with mercury searching for a permanent electric dipole moment. HFP is hexafluorobenzene.}
    \label{table:OMAGsParameters}
    \centering
    \begin{tabular}{c|c|c|c|c|c|c|c|c}
        \multirow{2}{*}{Name} & Element(s)/ & $\delta B_f$ & $\delta B_d$ & $\delta E_f$ & $\delta E_d$ & \multirow{2}{*}{$T_2$~[ms]} & \multirow{2}{*}{Spin} & \multirow{2}{*}{Ref.}\\
        & Compound(s) & $\left[\text{fT}/\sqrt{\text{Hz}}\right]$ & $\left[\text{fT}/\sqrt{\text{Hz}}\right]$ & $\left[10^{-20}\text{eV}/\sqrt{\text{Hz}}\right]$ & $\left[10^{-20}\text{eV}/\sqrt{\text{Hz}}\right]$ & & coupling & \\

        \toprule
        SERF & $^3$He & 0.002 & 0.75 & $3\times 10^{-5}$ & 0.01 & 10 & Nuclear & \cite{Vesilakis2009Limits,Kornack2002Dynamics}\\

        $\mu$-SERF & Rb & 1 & 30 & 1.9 & 58 & 10 & Total & \cite{Theis2012Zero}\\

        NMR-SERF hybrid & pentane-HFB & 0.23 & 3200 & 0.004 & 55 & $10000$ & Nuclear & \cite{Ledbetter2012Liquid}\\

        NMOR & Rb & 0.16 & 0.3\footnote{\label{footnote:1}The sensitivity was estimated based on the experimentally measured signal amplitude and width projected on the photon-shot-noise limited rotation for the used light intensity.} & 0.31 & 0.58 & 300 & Total & \cite{Budker2000Sensitive}\\

        AM NMOR & Rb & 3.2 & 39 & 9 & 110$^\text{\ref{footnote:1}}$ & 25 & Total & \cite{Pustelny2008Magnetometry}\\

        M$_x$ & Cs & 5 & 9 & 7 & 13 & 200 & Total & \cite{Castagna2009Large}\\

        $\mu$-M$_x$ & Cs & 20 & 42 & 29 & 61 & 0.06 & Total & \cite{Scholtes2011Light} \\

        Helium & He & 5 & 50 & 54 & 540 & 10000 & Electron & \cite{McGregor1987High} \\

        Hg EDM & Hg & $6\times 10^{-4}$\footnote{\label{footnote:2}This is an ultimate sensitivity limit calculated based a simplifying assumption of Eq.~(24) from Ref.~\cite{Swallows2013Techniques}.} & 320 & $2\times 10^{-6}$ & 1 & 100000 & Nuclear & \cite{Swallows2013Techniques}\\

    \end{tabular}
\end{table*}

In order to detect transient spin couplings, the signal characteristics must fall into the detection capabilities of the OMAGs used. Table~\ref{table:OMAGsParameters} summarizes characteristics of various OMAGs with potential applicability to the GNOME.

\section{Theoretical background\label{sec:Theory}}

A specific example of exotic spin coupling that may be detected with the GNOME is the transit of the Earth through a domain wall (DW) of a light pseudoscalar (axion-like) field \cite{Pospelov2013Detecting}. Stable domain structure of axion-like fields is a consequence of certain Standard Model extensions \cite{Dobrescu2006Spin,Sikivie1983Experimental,Friedland2003Domain,Jaeckel2010Low}. Domains form out of the initially random distribution of the vacuum expectation values of the axion-like field as the Universe expands and cools. In this scenario, DWs separate regions of space with different energy vacua \cite{Sikivie1983Axions}. Importantly, based on astrophysical constraints, only light axion-like fields can build DWs that persist to the present epoch \footnote{Formation of the wall from QCD-axion field would lead to disastrous cosmological consequences due to the excessive energy stored in the walls.}.

A detailed theoretical background of the optical detection of wall crossings is presented elsewhere \cite{Pospelov2013Detecting}. Here the concept is only briefly reviewed. We start with considering a hypothetical pseudoscalar field $a(\vect{r})$ that permeates the Universe and forms a domain structure. As shown in Ref.~\cite{Pospelov2013Detecting}, a specific realization of the field existing between neighboring domains with different energy vacua (with the DW centered at $z=0$)
\begin{equation}
    a(z)=4 a_0 \arctan[\exp(m_azc/\hbar)],
\label{eq:AxionField}
\end{equation}
where $a_0$ is the characteristic amplitude of the field and $m_a$ is the pseudoscalar-particle mass. Coupling between the axion-like-field gradient $\vect{\nabla}a$ and the spin $\vect{F}$ arising during the domain-wall crossing is described by the Hamiltonian \footnote{Note that in general, the higher-order couplings to the spins may be also considered (see Ref~\cite{Pospelov2013Detecting} for further details).}
\begin{equation}
    H_\text{DW}=\hbar c\frac{\vect{F}\cdot\vect{\nabla}a}{Ff_\text{eff}},
\label{eq:Interaction}
\end{equation}
where $f_\text{eff}$ is the effective decay constant in units of energy. $f_\text{eff}$ depends on the atomic structure of the particles used in a specific OMAG and is a combination of electron $f_\text{e}$, proton $f_\text{p}$, and neutron decay constants $f_\text{n}$. By substituting Eq.~(\ref{eq:AxionField}) into Eq.~(\ref{eq:Interaction}), the Hamiltonian $H_\text{DW}$ can be expressed using the field parameters $m_a$ and $a_0$
\begin{equation}
    H_\text{DW}=\frac{2c^2}{f_\text{eff}}\frac{a_0m_a\cos\varphi}{\cosh(m_az)},
\label{eq:InteractionSubstituted}
\end{equation}
where $\varphi$ is the angle between the spin $\vect{F}$ and the field gradient $\vect{\nabla}a$.

The thickness of the DW $d$ is determined by the pseudoscalar-particle mass $m_a$ via
\begin{equation}
    d=\frac{2\hbar}{m_ac}.
    \label{eq:DWthickness}
\end{equation}
Consequently, the mass also limits the duration of the transient signal $\Delta t=d/v_\bot$, where $v_\bot$ is the relative speed between the DW and the OMAG. A wall may be characterized by the average tension $\sigma$, which is the axion-like particle energy per unit area. In the considered case, it can be written as a function of the field parameters
\begin{equation}
    \sigma=\frac{4m_aa_0^2}{\hbar^2}.
    \label{eq:Tension}
\end{equation}
The tension can be related to the energy density of the DW network $\rho_\text{DW}$ via $\rho_\text{DW}\approx\sigma/L$, where $L$ is the characteristic size of the domain. Importantly, the density $\rho_\text{DW}$ needs to be smaller than the DM density $\rho_\text{DM}$ ($\rho_\text{DM} \approx 0.4$~GeV/cm$^3$) or the DE energy density $\rho_\text{DE}$ ($\rho_\text{DE} \approx 0.4 \times 10^{-5}$~GeV/cm$^3$). Determination of the tension also requires knowledge about the characteristic size of the domain $L$. Since it is not possible to determine $L$ without further assumptions about the specific mechanism of domain-structure formation, here one treats $L$ as a free parameter and constrain it from an experimental perspective, i.e., the experimental feasibility implies that the average time $T$ between two wall crossings should not be longer than 10 years. By taking into account the speed of the solar system relative to the Galactic frame ($v \approx 10^{-3}c$), a DW-crossing event will likely occur within a time-span of 10 years if the domain size is less than $10^{-2}$~ly.

Combining Eqs.~(\ref{eq:InteractionSubstituted}) and (\ref{eq:Tension}), one may obtain the experimental limit $f_\text{exp}$ on the effective decay constant ($f_\text{eff}\leq f_\text{exp}$)
\begin{equation}
    f_\text{exp}=\hbar c^2\frac{\sqrt{\rho_\text{DW}Lm_a}}{\delta E_d}\cos\varphi,
    \label{eq:FinalCoupling}
\end{equation}
where $\delta E_d$ is the demonstrated magnetometric sensitivity.

\begin{figure}[h!]
    \includegraphics[width=\columnwidth]{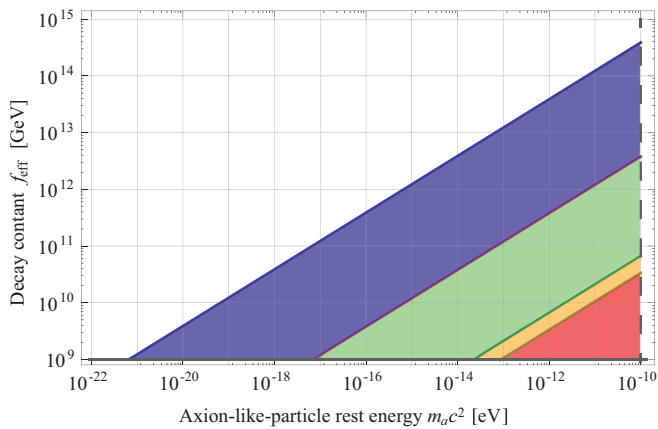}
    \caption{Parameter space of the axion-like field with a domain structure that can be probed with the GNOME for various DW energy densities. The vertical line at 10$^-10$~eV is set by the bandwidth of the measurements. The horizontal line at 10$^9$~GeV corresponds to the lower bound on electron, neutron, and proton decay constants. The diagonal lines represent saturation of the DW density ($\rho_{DW} =\rho_{DM}$) for various magnetometers. The vertical dashed line at $10^{-10}$~eV marks the estimated (rather than sharp) boundary arising from the assumed bandwidth of the magnetometer (100 Hz). The lines/shades correspond to the demonstrated sensitivity of the devices $\delta E_d$ and a measurement bandwidth of 100 Hz: blue - SERF magnetometer ($1\times 10^{-21}$~eV), purple - Hg magnetometer ($31\times 10^{-19}$~eV), yellow - $\mu$-SERF ($58\times 10^{-19}$~eV), and red - AM NMOR ($11\times 10^{-18}$~eV).}
    \label{fig:GNOME-constraints}
\end{figure}

Figure~\ref{fig:GNOME-constraints} shows the parameter space that can be probed with various OMAGs presented in Table~\ref{table:OMAGsParameters}.
The diagonal lines correspond to the limits due magnetic-field sensitivities of the devices [Eq.~(\ref{eq:FinalCoupling})] with saturated condition for the DW density ($\rho_\text{DW}=\rho_\text{DM}$) and a characteristic DW-size of $10^{-2}$~ly (10-year measurement). The horizontal line at 10$^9$~GeV is the existing lower bound on the electron $f_e$, neutron $f_n$, and $f_p$ proton decay constants due to astrophysical constraints. The vertical line at the right-hand side of the plot is determined by the bandwidth of the measurements. Here, for simplicity, it is assumed that all magnetometers have a 100 Hz bandwidth corresponding well to the bandwidth of the magnetometers employed in our demonstration experiment (see next section). As shown, the technique allows probing a significant part of presently unconstrain parameter space of the axion-like field. It should be also noted that for low $m_a$ the thickness of the wall becomes large so that the transition through the DW looses its transient character and becomes a quasi-stationary process.

In general, the range of axion-like-particle mass that can be probed with an OMAG may be extended by increasing the bandwidth of the devices. While this causes linear increase in a mass range of particles probed, the sensitivity of the measurement is reduced, which shifts the diagonal limit down by the square-root of the bandwidth. This allows to improve the size of the probed parameter space, which may be interesting in various scenarios.

\begin{figure}[h!]
    \includegraphics[width=\columnwidth]{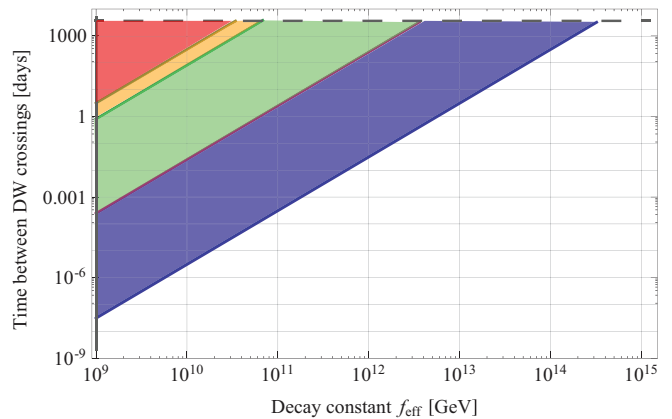}
    \caption{Average time between successive DW crossings $T$ as a function of the effective decay constant $f_\text{eff}$ of an axion-like-particle energy of 10$^{-10}$~eV. The dashed horizontal line correspond to the 10-year envision period for the experiment, while the solid vertical line is due to astrophysical constrains. The color code is the same as in Fig.~\ref{fig:GNOME-constraints}.}
    \label{fig:GNOME-time}
\end{figure}

Detection of DW crossing of the axion-like field characterized with a decay constant within the range presented in Fig.~\ref{fig:GNOME-constraints} requires an ability to record a short transient event ($\gtrsim 10$ ms) within a 10-year time span. Reliable operation of an OMAG at the top of its performance for such a long time, is an extremely challenging task. Therefore, it is instructive to investigate a characteristic time $T$ between two wall crossings, and hence anticipated problem-free operation time of the device, versus the axion-like-field parameters. Figure~\ref{fig:GNOME-time} presents the time as a function of the decay constant $f_\text{eff}$ for the OMAGs shown in Table~\ref{table:OMAGsParameters}.
The results show that an OMAG can probe significant regions of parameter space in less than a week of continuous operation and this region expands rapidly with increasing duration of the measurements.

In principle, the parameters of the model can be constrained with a single magnetometer. A particular problem for a search carried out with a single OMAG is the appearance of brief spikes in the OMAG signal related to technical noise or abrupt magnetic field changes. In a single device, rejection of these false-positive signals is difficult. At the same time, coincident measurements between two or more instruments are helpful in rejecting such signals; they provide consistency checks, since a signal would be expected to exist in all instruments whereas environmentally induced events are not typically correlated in the time window required for coincidence. Furthermore, information about a putative event such as its impinging direction can be determined by triangulation if several instruments (at least four) are taking data simultaneously (see discussion in Sec.~\ref{sec:Prospects}). These features clearly show that synchronous operation of multiple synchronized, geographically separated OMAGs within the proposed global network may facilitate searches for such transient signals of astrophysical origin.

\section{Experimental apparatus\label{sec:Setup}}

\begin{figure}
    \includegraphics[width=10cm, trim=0cm 2cm 0cm 5 cm]{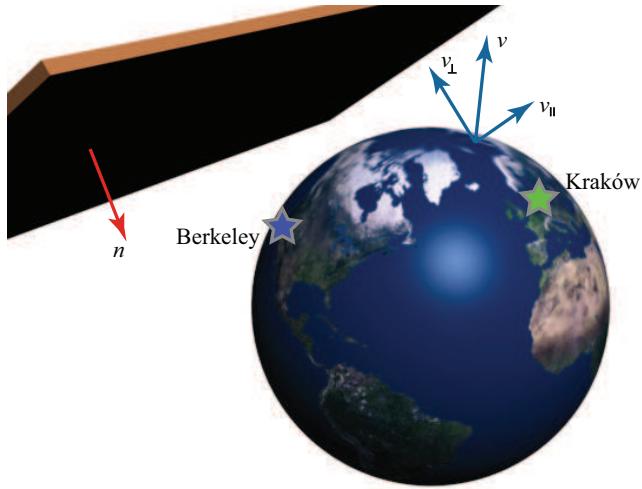}
    \caption{The concept of the synchronized-magnetometer arrangement. OMAGs located at globally separated locations record signals with a time synchronization provided by the GPS. By synchronously detecting and correlating magnetometer signals, transient events of global character may be identified. In particular, correlating signals of at least four devices enables detection of spatiotemporal character of the event. Here, two devices located in Krak\'ow and Berkeley are used to search for transient signals induced by crossing of a DW of an axion-like field (surface at the upper left of the figure). Blue arrows indicates the Earth velocity and velocity components with respect to the normal to the wall (red arrow).\label{fig:setup}}
\end{figure}

The concept of the experimental apparatus is shown in Fig.~\ref{fig:setup}.
Both magnetometers use rubidium vapors as the magneto-optically active medium. In the Krak\'ow magnetometer, the atomic vapor is contained in a paraffin-coated evacuated cylindrical glass cell with volume $\approx 3$~cm$^3$. The vapor cell is maintained at about $50^\circ$C corresponding to an atomic density of roughly 10$^{11}$~atoms/cm$^3$. The relaxation rate of the atomic ground state is 2$\pi\times 30$~s$^{-1}$, which yields a fundamental sensitivity $\delta B_f$ of $\approx3$~fT/$\sqrt{\text{Hz}}$ (spin-projection limited) \cite{Pustelny2008Magnetometry}. The second magnetometer (Berkeley) exploits a microfabricated vapor cell \cite{Schwindt2004Chip} of a volume of 0.01~cm$^3$ that is heated up to about $200^\circ$C. Operation in the spin-exchange relaxation free regime \cite{Allred2002High} allows elimination of relaxation due to spin-exchange collisions, one of the main ground-state polarization-relaxation mechanisms. Application of the technique allows one to obtain a ground-state relaxation rate of about $2\pi\times 10$~s$^{-1}$, which in combination with 3-4 orders of magnitude higher density yields a similar sensitivity as for the other setup ($\sim$1~fT/$\sqrt{\text{Hz}}$) \cite{Theis2012Zero}. Both magnetometers are thus capable of detecting a DW crossing and probing the parameter space.

Both magnetometers are placed inside multilayer magnetic shields made of $\mu$-metal with the innermost layer made of ferrite \footnote{While magnetic shields do not screen exotic interactions, their role in searches for transient exotic couplings require more thorough investigations in the future.}. The shields reduce external magnetic fields by a factor 10$^6$. Inside the shield atoms are subjected to a stable, well-controlled magnetic field generated by a set of three-dimensional magnetic-field coils. In the Krak\'ow magnetometer a field with a magnitude of 10$^{-7}$~T is applied, while at Berkeley the applied-field magnitude is $\approx5\times 10^{-8}$~T.

The outputs of the magnetometers are acquired using custom-made devices based on Trimble Resolution-T GPS (Global Positioning System) time receivers \cite{Wlodarczyk2013GPS}. The data acquisition devices provides time markers separated by one second with a precision of about 80~ns synchronized with a quartz clock built into the devices. The acquisition devices can record simultaneously signals in four channels at a rate of 1000 samples/s. Each one-second-long record is stored on a memory card with a header containing information on time, measurement condition, GPS-device warnings, etc. The records are transmitted to a computer (via serial port) where they are binned into groups of 10-1000 (typically 2-minute long bins are generated). The data are stored with computers located at the respective locations, and every 1-2 hours the information is exchanged between Krak\'ow and Berkeley using File Transfer Protocol (FTP). In this manner, the complete set of data is accessible at both locations.

\section{Results and Data Analysis\label{sec:Results}}

\begin{figure}
     \includegraphics[width=\columnwidth]{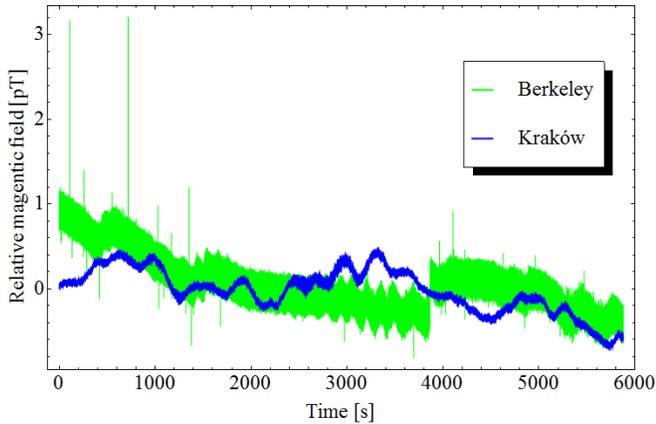}
     \caption{Synchronously detected magnetic fields measured with the OMAGs located in Krak\'ow and Berkeley. The signal in the Krak\'ow magnetometer was measured at a field of 100~nT, whereas the Berkeley magnetometer operated at 50~nT (DC offsets are removed from the plots). Note that the DC magnetic field in the Berkeley experiment was modulated with an AC field of a magnitude of $5\times 10^{-12}$~T oscillating at a frequency of about $2$~Hz. In the measurements, the data acquisition rate was 256 samples per second.\label{fig:OMAGdata}}
\end{figure}

Figure~\ref{fig:OMAGdata} presents magnetometer signals measured synchronously at two locations (Berkeley, California, USA and Krak\'ow, Poland) over a period of about 1.5~hours.
The long-term drift of the Krak\'ow magnetometer is most likely attributable to instability of the laser frequency, which, in the particular arrangement, mimic magnetic-field changes. This problem will be addressed in the future by implementation of laser-stabilization techniques. At the same time, the drift of the Berkeley magnetometer is most likely induced by the instability of the magnetic field inside the shield and/or imperfections in shielding external fields. These drifts may be limited by either active compensation of external magnetic fields or by correlating magnetometer signal with readout of sensors situated outside the shield, e.g., magnetometers, thermometers, etc. The Berkeley magnetometer also exhibits short-duration ($\lesssim 4$~ms) spikes of relatively large amplitudes.  Auxiliary tests verified that these noise spikes originated from electronic pick-up, a problem that will be addressed in the future.

In many respects, the identification of a DW-crossing event using the GNOME is similar to searches for gravitational-wave bursts with a system of long-baseline laser interferometers such as the Laser Interferometer Gravitational Wave Observatory (LIGO), the Virgo detector, GEO 600, and TAMA 300 \cite{Abadie2012All}. Both types of experiments aim to identify and characterize transient signals and search for time-domain correlations between the transient signals measured with different detectors. Importantly, the field of transient gravitational-wave astronomy has developed a variety of statistical methods to identify brief (duration $\lesssim 1$~s) signals correlated among different detectors but otherwise generic in noisy time series data.

As a proof-of-principle demonstration, we have applied one of the methods upon which such statistical analysis is based, the ``excess power'' statistic \cite{Anderson2013Excess}, to the synchronous magnetometer data from the Krak\'ow and Berkeley sites.  The analysis is carried out as follows.  First, an estimation of the power spectral density (PSD) over several continuous, overlapping segments of the data is made for an individual OMAG. These spectra are calculated at regular intervals and combined with previous measurements using a running median exponentially-weighted history (this is a filter that weighs contributions to the moving average by a factor that decreases exponentially with the time since the data were acquired). Data are then normalized by the corresponding PSD bin, which flattens the data spectral profile, ideally, producing a stream of Gaussian distributed, zero-mean, unity variance random variables characterizing the data set (data whitening). This allows obtaining data of similar spectral properties independently from the characteristics of the actual magnetometer used for the measurements. The stream of whitened data is then passed through a bank of band-limited filters producing several channels of filtered data. These filters are Hann windows with a width corresponding to the bandwidth of each channel $\delta\nu_f$ and centered at a set of frequencies separated by the bandwidth. The set of filters is constructed so that it completely spans the entire bandwidth of the input data. The filters themselves are also normalized by the PSD in the frequency domain, i.e., they are ``whitened'', to better predict the response of a signal in a given channel. Temporally adjacent data samples from each channel are summed to form a discrete localization of energy in the original data stream in frequency and time. At these locations, the, so-called, ``tiles'' are formed, whose time-frequency bounds are determined by the bandwidth of the filters $\delta\nu_f$ and duration $\delta t$ corresponding to the number of summed samples. The resulting tile has $N_\text{DOF}$ independent degrees of freedom, which are determined by the product of the bandwidth $\delta\nu_f$ and time $\delta t$, $N_\text{DOF}=2\delta_t \delta\nu_f$.

\begin{figure}[htbp]
     \includegraphics[trim = 20mm 30mm 10mm 0mm,width=\columnwidth]{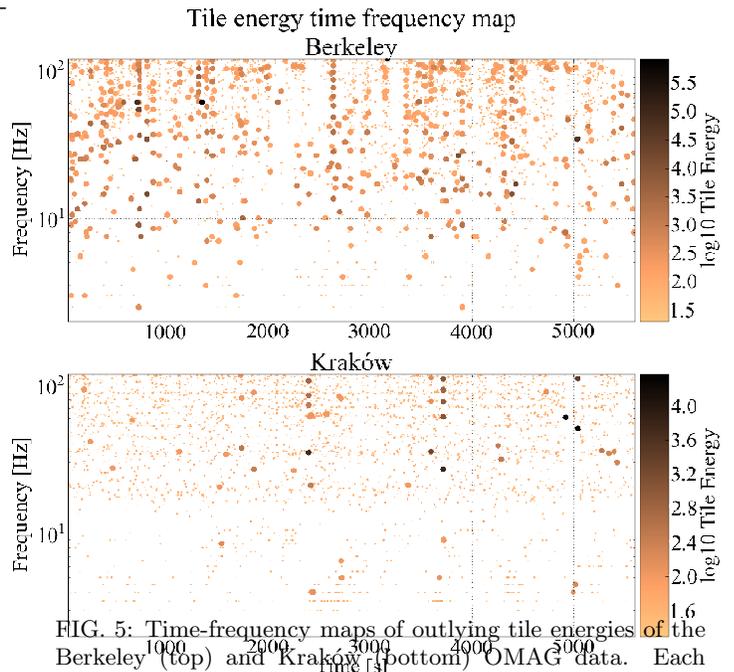}
     \caption{Time-frequency maps of outlying tile energies of the Berkeley (top) and Krak\'ow (bottom) OMAG data. Each dot represents a tile characterized by its power-weighted central time ($x$-axis), Fourier frequency ($y$-axis), and tile energy (color and dot size). The large markers indicate tiles with a normalized tile energy greater than 100. Large tile energies are likely caused by discontinuities (which cause short but broadband responses) in the data or environmental influences within the instruments themselves.\label{fig:trigmap}}
 \end{figure}

The final product is a time-frequency map of tiles describing the whitened-signal energy. Under the conditions of stationarity (the PSD does not fluctuate on the time scale of the estimation process) and Gaussianity (the data samples have a distribution matching the Gaussian distribution), the tile energies are distributed as a $\chi^2$ distribution with $N_\text{DOF}$ degrees of freedom. Thus the significance of any tile's energy is well-understood and the statistical probability of outlier tiles can be measured.

\begin{figure}
     \includegraphics[width=\columnwidth]{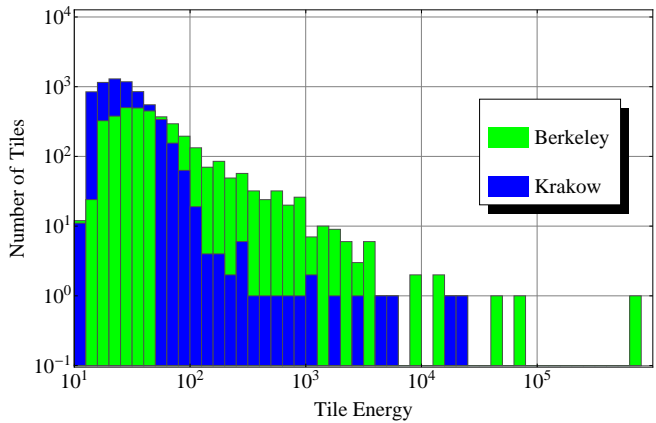}
     \caption{Histogram showing the number of tiles with energy within a energy range for the magnetometer data depicted in Figs.~\ref{fig:OMAGdata} and \ref{fig:trigmap}.}
     \label{fig:histogram}
 \end{figure}

Figure~\ref{fig:trigmap} presents the time-frequency maps of the data presented in Fig.~\ref{fig:OMAGdata}. Examination of the map reveals that the Krak\'ow instrument produces fewer ``loud'' transients (associated with high tile energies) than the Berkeley instrument, which indicates that the Krak\'ow data more closely follow the signal expected from Gaussian-distributed noise.  This is more visible in the histogram of the tile energies shown in Fig.~\ref{fig:histogram}. The observed roughly one order of magnitude excess of events in the Berkeley magnetometer signal compared to the Krak\'ow signal for tile energies above $\gtrsim 100$ indicates problems in the relative quality of the Berkeley data compared to the Krak\'ow data, manifesting in the raw data (Fig.~\ref{fig:OMAGdata}) by appearance of the spikes.

Correlated transient signals from different magnetometers can be searched for by using the time-frequency tile maps (Fig.~\ref{fig:trigmap}) to find overlapping events with related characteristics. Time shifts can be introduced into the data to check for correlated transient events with particular relative delays. These delays are determined by the time of DW travel between two magnetometer, i.e., the projection of the wall velocity on the baseline of the OMAGs. In the most extreme case $v_\bot=10^{-3}c$ and the delay time between two triggers spans between $-30$~s to 30~s. For the purpose of this analysis, it is assumed that DW crossing occurs simultaneously in both magnetometers, so that there is no delay between the appearance of the transients in Krak\'ow and Berkeley time-frequency maps. In the future implementation of the analysis, however, other time delays will be also considered to accommodate for other scenarios, i.e., to include various delays between the signals and durations of the wall crossings.

Existence of the time window within which the DW-crossing event is detected by the OMAGs enables determination of a background of false-positive events. The window duration is determined by the wall thickness (originating from the axion-like-particle mass $m_a$) and a maximum possible delay between the triggers (arising from the speed of Earth normal to wall $v_\bot$), which is assumed here to be zero. For the delays larger than the window, correlation of the signals provide a false-positive background of the measurements and hence the threshold for the real events.

\begin{figure}
    \includegraphics[width=\columnwidth]{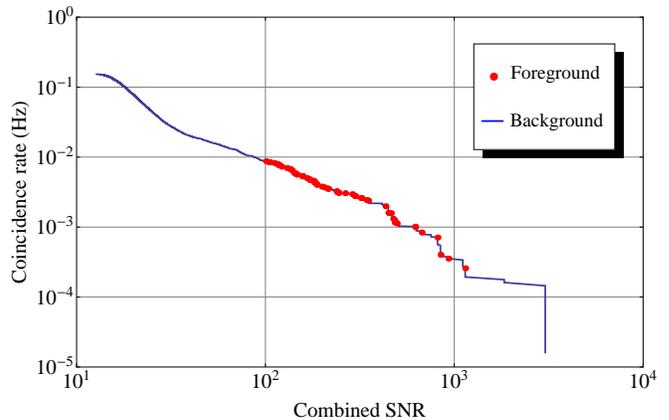}
    \caption{The cumulative rate of false positives as estimated from the shifting procedure. Only tiles with duration of less than 2~s are used to form coincidences, since this corresponds to an upper limit on how long the signal is expected to reside within a magnetometer. Ten unique offsets in multiples of 100 seconds were used to build the curve. The foreground of putative events is plotted with red dots overlaid on the curve to indicate the the measured rate of false positive at that level of combined SNR. Only events with a combined SNR of greater than 100 are shown.\label{fig:far_curve}}
\end{figure}

Figure \ref{fig:far_curve} shows the cumulative rate of coincidence versus their combined SNR for the background of false positives.
Event candidates obtained from the correlation of non-offset signals are overlayed on top of this curve to indicate the false-positive rate. The list of coincidences from the non-offset data is then compared to the estimated cumulative rate of false positives. At the SNR of the potential candidate, the rate defines how often one might expect a false-positive signal from the coincidence analysis. The statistical significance of a specific event is then derived from a Poissonian distribution, using the estimated false-positive rate and the observation time to form the rate constant.

The background can be further constrained by analysis of the signals from more OMAGs. In such a case, the multiparty coincidence would reduce the background and put further constrains on the real signal. Moreover, with at least four magnetometers triangulation may be performed (velocity $v_\bot$ may be determined), so that predictions for additional devices may be made. This would work as a mechanism for further elimination of the false-positive signal.

\section{Prospects\label{sec:Prospects}}

Comagnetometry, where the magnetic field is simultaneously measured with multiple atomic species or devices, is a widely used technique in precision measurements searching for anomalous spin-dependent effects (see, for example, Ref.~\cite{Budker2013Optical} and references therein). Comagnetometry with different atomic species takes advantage of the fact that the relative coupling strengths of an exotic field to electrons and nuclei are generally different from the relative coupling strengths of electrons and nuclei to magnetic fields. A particular example of a comagnetometry scheme that will be investigated for possible use as a GNOME sensor is SERF comagnetometer similar to that described in Ref.~\cite{Smiciklas20011New}. In contrast to the devices used in the demonstration experiment described in the present work, an additional noble gas (helium) is introduced into a vapor cell. When the noble gas has non-zero nuclear spin, the alkali and noble gas spins become strongly coupled through spin-exchange collisions \cite{Kornack2002Dynamics,Kornack2005Nuclear}. This coupling can be represented as the effective magnetic field $\vect{B}_\text{eff}$ experienced by one spin species due to the average magnetization $\vect{M}$ of the other, due to enhancement of the alkali valence electron density at the noble gas nucleus,
\begin{align}
\vect{B}_\text{eff} = \lambda \vect{M},
\end{align}
where $\lambda$ is a parameter determined by the particular properties of the alkali-noble gas spin-exchange \cite{Walker1989}. The applied field $\vect{B}$ is tuned so that it approximately cancels $\vect{B}_\text{eff}$ experienced by the alkali atoms.  The alkali atoms are then in an effective zero-field environment, and because the noble gas magnetization $\vect{M}$ adiabatically follows $\vect{B}$, transverse components of $\vect{B}$ are automatically compensated by $\vect{B}_\text{eff}$ to first order. Such cancelation only occurs for interactions that couple to spins in proportion to their magnetic moments, leaving the SERF comagnetometer sensitive to anomalous spin couplings to electrons and nuclei \cite{Kornack2002Dynamics}.

The response of a SERF comagnetometer to a transient event such as a DW crossing can be understood based on the work described in Ref.~\cite{Kornack2002Dynamics}, where the spin dynamics of a $^{3}$He-K SERF comagnetometer were studied.  A DW crossing event would generate brief torques, nominally of different magnitudes, on the $^{3}$He and K spins The resulting hybrid oscillatory response of the spin ensemble would decay at approximately the electron spin relaxation rate. The advantage of a SERF comagnetometer is its ability to self-compensate environmental magnetic fields and detect transient events even if coupling to electron spins may be reduced/screened by a magnetic shield.

The energy resolution of the latest generation of the SERF comagnetometer, employing Rb as the alkali atom and $^{21}$Ne as the noble gas, is $\sim 10^{-23}~{\rm eV/\sqrt{Hz}}$ \cite{Smiciklas20011New}.  This new scheme uses hybrid optical pumping of Rb via spin-exchange collisions with low-density, optically pumped K and off-resonant direct optical probing of Rb spins. This approach allows full optimization of both optical pumping and probing.  Because of the relatively small gyromagnetic ratio of $^{21}$Ne, the Rb-K-$^{21}$Ne SERF comagnetometer has an order of magnitude better energy resolution for the same level of magnetic-field sensitivity as compared to earlier SERF comagnetometers, and may offer advantages in bandwidth. In the future, it is planned to develop and optimize the SERF-based comagnetometer for measurements of exotic transient effects.

Independently from the development of SERF-based comagnetometer, the other magnetometer types will be developed as potential GNOME sensors. In particular, it is envisioned using sensors that monitor evolution of various types of spins (proton, neutron, electron). This would add another dimension to our investigations by studying influence of exotic coupling to various fundamental particles.

Another important work envisioned for a future experiment is correlation of the magnetometer readouts with environmental parameters (e.g., magnetic field outside the shield, temperature, etc.). This is motivated by the fact that despite magnetic shielding, there will inevitably be some level of transient signals and noise associated with the local environment (and possibly with global effects like the solar wind, changes of the Earth's magnetic field, etc.). The environmental-condition data will allow for exclusion/vetoing of data with known systematic issues.

A further step in reducing the influence of magnetic fields on the operation of GNOME is application of Superconducting Quantum Interference Device (SQUID) \cite{Clarke2004SQUID} magnetometers as sensors operating in addition to OMAGs inside the magnetic shields. While the SQUID magnetometers are characterized with magnetometric sensitivity comparable to that of OMAGs, they are not sensitive to exotic spin coupling. Thus, they can be used for vetoing false-positive transient signals.

Ultimately, the GNOME will consist of at least five OMAGs. Four devices will be used for the detection of a DW and of its geometrical properties. Any additional magnetometer would increase the sensitivity of the network. An independent OMAG will serve as cross-check to verify if, based on predicated DW event, a transient signal arises in the magnetometer in a narrow temporal window.

\section{Conclusions\label{sec:Conclusions}}

In this paper, a new experimental scheme enabling investigations of transient exotic spin couplings has been presented. It is based on synchronous operation of globally separated optical magnetometers enclosed inside magnetic shields. Correlation of magnetometers' readouts enables filtering local signals induced by environmental and/or technical noise. Moreover, application of vetoing techniques, e.g., via correlation of optical-magnetometer readouts with signals detected with non-optical magnetic-field sensors, enables suppression of influence of global disturbances of magnetic origins, such as solar wind, fluctuation of the Earth's magnetic field, on the operation of the magnetometers. In such an arrangement, the network becomes primarily sensitive to spin coupling of non-magnetic origins, thus it may be used for searches of physics beyond the Standard Model. A specific example of such searches was discussed here by considering coupling of atomic spins to domain walls of axion-like fields. It was demonstrated that with modern state-of-the-art optical magnetometers probing a significant region of currently unconstrained space of parameters of the fields is feasible. The preliminary results obtained based on synchronous operation of two magnetometers located in Krak\'ow and Berkeley were presented and future plans for the network development were outlined.

\begin{acknowledgments}
The authors are thankful to S. Bale, J. Clarke, S. Rajendran, A. Sushkov, and M. Zolotorev for useful discussions. S.P. is a scholar of the Polish Ministry of Science and Higher Education within the Mobility Plus Program. D.B. acknowledges the support by the Miller Institute for Basic Research in Science. This work has been supported in part by the National Science Foundation under grants: PHY-0969666, PHY-1068875, PHY-0970074, PHY-0970147, and the ``Team'' Program of the Foundation for the Polish Science.
\end{acknowledgments}

\end{document}